\documentclass[10pt]{ieeetran}

\usepackage[left=54pt, right=54pt, bottom=54pt, top=72pt]{geometry}
\title{On SORA for High-Risk UAV Operations under New EU Regulations: Perspectives for Automated Approach 
}
\author{Hamed Habibi \textit{Member, IEEE}, D. M. K. K. Venkateswara Rao, Jose Luis Sanchez-Lopez, and Holger Voos
\thanks{This work is partially funded by the Department of Media, Telecommunications and Digital Policy (SMC) of the Government of the Gran Duchy of Luxembourg under the project reference SMC/CFP-2019/010/IRANATA, "Interference and RAdiation in Network PlAnning of 5G AcTive Antenna Systems."}%and by the European Union’s Horizon 2020 project Secure and Safe Multi-Robot Systems (SESAME) under the grant agreement no. 101017258."}
 \thanks{H. Habibi, D. M. K. K. Venkateswara Rao, J. L. Sanchez-Lopez, and H. Voos are with Automation and Robotics Research Group, Interdisciplinary Centre for Security, Reliability and Trust, University of Luxembourg, Luxembourg. H. Voos is also with Faculty of Science, Technology and Medicine (FSTM), Department of Engineering, University of Luxembourg, 
{\tt\small \{hamed.habibi,mohan.dasari,holger.voos,\\joseluis.sanchezlopez\}@uni.lu}
}
}
\date{}
\usepackage{amsthm}
\usepackage{amssymb}
\usepackage{amsmath}
\usepackage{breqn}
\usepackage{enumitem}
\usepackage{hyperref}
\usepackage{tikz}
\hypersetup{
    colorlinks=true,
    linkcolor=blue,
    citecolor=red,
}
\usepackage{color,soul}
\usepackage{graphics} 
\usepackage{epsfig}
\usepackage{mathptmx} 
\usepackage{times}
\usepackage{amssymb}
\usepackage[mathscr]{euscript}
\usepackage{enumerate}
\usepackage{algpseudocode}
\usepackage{algorithm}
\usepackage{physics}
\usepackage{multirow}
\usepackage{optidef}

\newtheorem{remark}{Remark}

\DeclareMathAlphabet{\mbf}{OT1}{ptm}{b}{n}

\graphicspath{{figures/}}

\begin{document}
% \author{\IEEEauthorblockN{D. M. K. K. Venkateswara Rao, Hamed Habibi, Jose Luis Sanchez-Lopez, Holger Voos}
% \IEEEauthorblockA{\textit{Interdisciplinary Centre for Security, Reliability and Trust} \\
% \textit{University of Luxembourg}\\
% Luxembourg, Luxembourg \\
% \{mohan.dasari, hamed.habibi, joseluis.sanchezlopez, holger.voos\}@uni.lu}
% }
\maketitle
\begin{abstract}
% In this paper, we investigate the required steps to prepare the application for Specific Operations Risk Assessment (SORA) to obtain flight authorization for outdoor drone operations. Even though the preparation of this application might seem as a governmental bureaucracy, it contains technicalities that require expert knowledge. Furthermore, the whole process is an iterative and time-consuming one, as we have observed throughout our experience, which is briefly presented. Accordingly, we propose an alternative workflow, addressing the observed challenges and pitfalls, to shorten the whole process. Furthermore, we present a comprehensive list of preliminary procedures to be prepared, including the devised pre/during/post-flight checklists, design and installation appraisal, flight logbook, operational manual, training manual, General Data Protection Regulation (GDPR) procedures and Training Records, which are not explicitly instructed in the manual. More importantly, we suggest an envisioned automated workflow, which can be implemented as a computer program, to be followed by the applicant. The presented approach can help drone operators, especially the research centres, since the need to obtain this authorization has been mandated in the last few years by corresponding regulatory bodies, which impose difficulty and delays to conduct experimental operations. 
%%%%%%%%%%%%%%%%%%%%%%%%%%%%%%%%%%%%%%%%%%%%%%%%%%%%%%%%%%%%%%%%%%%%%
In this paper, we investigate requirements to prepare an application for Specific Operations Risk Assessment (SORA), regulated by European Union Aviation Safety Agency (EASA) to obtain flight authorization for Unmanned Aerial Vehicles (UAVs) operations and propose some perspectives to automate the approach based on our successful application. Preparation of SORA requires expert knowledge as it contains technicalities. Also, the whole process is an iterative and time-consuming one. It is even more challenging for higher-risk operations, such as those in urban environments, near airports, and multi- and customized models for research activities. SORA process limits the potential socio-economic impacts of innovative UAV capabilities. Therefore, in this paper, we present a SORA example, review the steps and highlight challenges. Accordingly, we propose an alternative workflow, considering the same steps, while addressing the challenges and pitfalls, to shorten the whole process. Furthermore, we present a comprehensive list of preliminary technical procedures, including the pre/during/post-flight checklists, design and installation appraisal, flight logbook, operational manual, training manual, and General Data Protection Regulation (GDPR), which are not explicitly instructed in SORA manual. Moreover, we propose the initial idea to create an automated SORA workflow to facilitate obtaining authorization, which is significantly helpful for operators, especially the scientific community, to conduct experimental operations.
\end{abstract}

\section{Introduction} \label{sec:introduction}
In the last decade, there has been a significant increase in the use of Unmanned Aerial Vehicles (UAVs) in outdoor environments for different civilian applications, both in industry and academia \cite{srivastava2020review}. Moreover, there is a constant growth in the size, speed and maneuverability of modern UAVs. Consequently, in the last few years, governments have started regulating UAVs outdoor flights, to guarantee the safety and security of operations, concerning the people, other manned air traffic, and the environment \cite{dronelaws}. For instance, there have been already regulations in place in European Union (EU) \cite{EASA}, namely, Specific Operations Risk Assessment (SORA) methodology, USA \cite{usalaws}, China \cite{chinalaws}, and Canada \cite{canadalaws}, just to mention few.

These new regulations force the UAV operators to initially obtain authorization from the regulating body \cite{srivastava2020review,xu2020recent}. This need might cause some delays in conducting operations, which is crucial, specifically, for the research community, which does a wide range of experimental operations \cite{stocker2017review}. This stems from the hard and time-consuming process of preparing the application and, in most cases, the process is difficult to interpret and hence, iterative. Furthermore, the whole process requires expert knowledge, as it contains technicalities. This implies the need to revisit and investigate the whole process from a technical point of view to modify and shorten the workflow, in a way to avoid delays.

Motivated by this consideration, in this paper we investigate the whole SORA methodology, proposed by European Union Aviation Safety Agency (EASA) \cite{EASA}. Moreover, we briefly present our own application, in which we identify the methodology challenges and drawbacks. To the best of our knowledge, it is the first time the SORA workflow is investigated technically. Consequently, we present, generically, preliminary steps and documents to be prepared by the applicant to address identified challenges that facilitate the whole process. On this basis, we propose an alternative workflow, through which the application can be made in a straightforward manner with fewer iterations. More importantly, we envision a fully/semi-automated approach, in which most of the steps and criteria can be implemented as an automatic computer software.

The rest of the paper is organized as follows. In Section \ref{sec:methodology}, we review SORA methodology. In Section \ref{sec:aplication}, we present our own application of this methodology. In Section \ref{discussion}, we highlight the challenges in the current workflow and present a solution for the identified issues. Furthermore, we suggest an alternative workflow as well as an automated approach. The concluding remarks are given in Section \ref{sec:conclusions}.

\section{SORA Methodology} \label{sec:methodology}
\begin{figure*}[!htpb]
\centering
\resizebox{.6\linewidth}{!}{\input{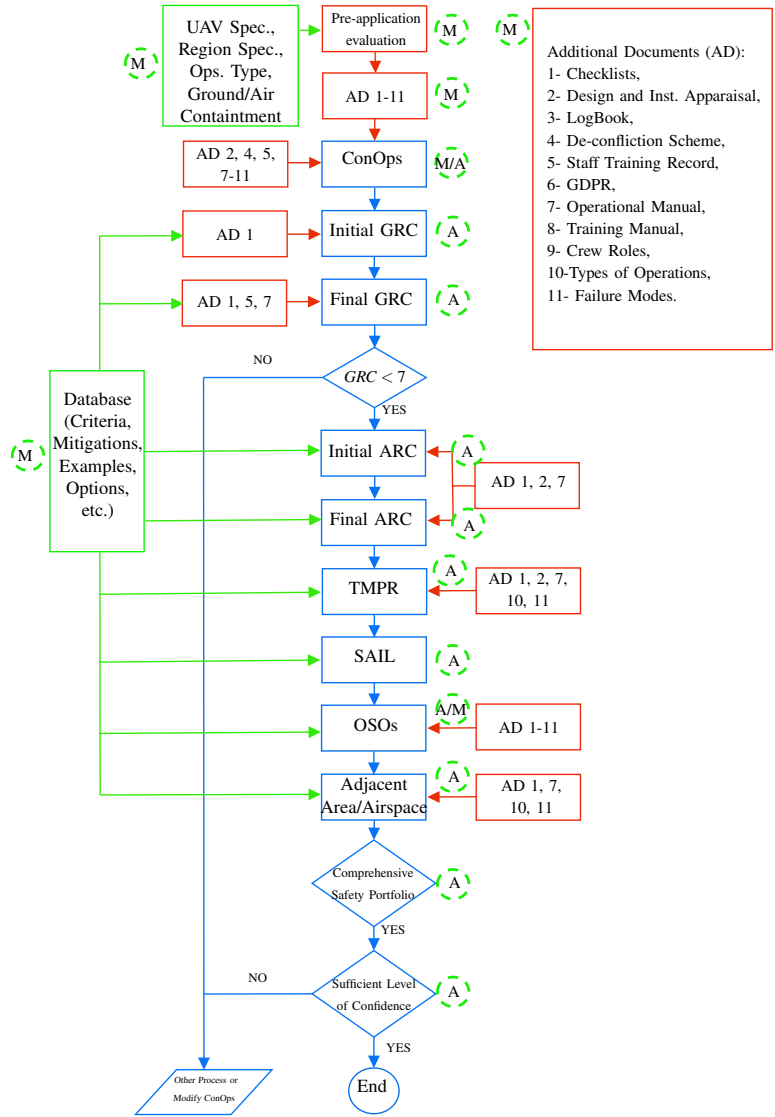}}
\caption{SORA methodology flowchart (blue boxes and lines). The proposed alternative approach to implementing the identified additional documents is in red. The automated approach is in green. The green dashed circles show the steps that can be automated (A) and manual (M). For ConOps is it M/A, $i.e.$, the first part is manual and the second part is automated. For OSOs it is A/M, $i.e.$, the criteria are obtained automatically, while the mitigations are manually selected.}
\label{fig:flowchart}
\end{figure*}

SORA methodology is a 10-step procedure, proposed by EASA, for applicants to carry out a risk assessment and obtain flight authorization, by addressing mitigations. SORA methodology, shown in Fig. ~\ref{fig:flowchart}, is described as follows.

\textbf{0) Pre-application evaluation:} As identified by EASA, there are three UAV types, namely, open, specific, and certified categories. As instructed, the applicant should verify that the intended operation is feasible and falls under the specific category. Also, it is not subject to a specific No-Go from the competent authority. Moreover, the operation is neither covered by a ‘standard scenario’ of the UAV regulation nor by a predefined risk assessment published by EASA. In such a case, SORA should be applied.

\textbf{1) Concept of Operations (ConOps):}
ConOps is essentially the main part of the application, which requires presenting the relevant organizational, operational and technical information. It provides background for all the subsequent risk assessments and mitigations. 
ConOps, categorically, includes two parts. The first part presents general information about the operation and the organization. This consists of the introduction of the organization with safety management protocols. Furthermore, the UAV design and production are reported here. The competence of all the flight crew is identified as per the staff training records. Also, any pre/post-flight maintenance activities performed are identified. The crew and their roles are introduced. UAV configurations and changes are to be reported via, $e.g.$, a design and installation appraisal form. In terms of the operation, the types of operation are introduced with general conditions and limitations. Furthermore, the standard operating procedures are to be defined, including normal, contingency and emergency procedures, occurrence reporting procedures, and Emergency Response Plan (ERP). 

In the second  part, all the necessary and sufficient technical information of the UAVs and supporting systems are collected and provided to address the required robustness levels of the mitigations. Initially, UAV segments and characteristics are introduced, including airframe, propulsion system, flight control surfaces and actuators, sensors and payloads. On this basis, the control segment specifications are given, $i.e.$, the avionics architecture, navigation, autopilot, flight control system, remote pilot station, and Detect and Avoid (DAA) system. More importantly, in this part, the containment system is introduced to ensure the UAV is confined to fly in the chosen region of flight. Also, the Ground Support Equipment (GSE) and Command and Control (C2) link segments are identified. Finally, the overall safety features and procedures, specifically for C2 link degradation and C2 link loss, are given.

\textbf{2) Intrinsic Ground Risk Class (GRC):}
The intrinsic GRC shows the risk of a person getting struck by a UAV that is out of control. This class is determined based on the maximum characteristic dimension of the UAV, its expected kinetic energy, and its operational scenario as shown in Table~\ref{table:intrinsic_grc}. The maximum characteristic dimension could be the wingspan, blade diameter, or maximum dimension for a fixed-wing, rotorcraft, or multi-copter, respectively. The expected kinetic energy is calculated based on cruise and terminal velocities. The operational scenario is determined based on the type and area of operation. The type of operation could be Visual Line-Of-Sight (VLOS) or Beyond VLOS (BVLOS). The area of operation could be a controlled, sparsely populated, populated zone, or an assembly of people. Additionally, the operational volume composed of flight geography, contingency volume, and ground risk buffer should also be determined.

\begin{table*}[!ht]
\centering
\caption{Determination of the intrinsic GRC.}
\label{table:intrinsic_grc}
\begin{tabular}[!h]{ c c c c c }
\hline\hline
\multicolumn{5}{c}{\bf Intrinsic UAV GRC}\\
\hline
{\bf Max. UAV characteristics dimension} & {  1 $m$} & { 3 $m$} & { 8 $m$} & { $\geq$ 8 $m$}\\
{\bf Expected kinetic energy} & { $<$ 700 $J$} & { $<$ 34 $kJ$} & { $<$ 1084 $kJ$} & { $>$ 1084 $kJ$}\\
\hline
{\bf Operational scenarios}\\
\hline
VLOS/BVLOS over a CGA & 1 & 2 & 3 & 4\\
VLOS over a sparsely populated area & 2 & 3 & 4 & 5\\
BVLOS over a sparsely populated area & 3 & 4 & 5 & 6\\
VLOS over a populated area & 4 & 5 & 6 & 8\\
BVLOS over a populated area & 5 & 6 & 8 & 10\\
VLOS over an assembly of people & 7 & NA & NA & NA\\
BVLOS over an assembly of people &  8 & NA & NA & NA\\
\hline\hline
\end{tabular}
\end{table*}

\textbf{3) Final GRC:}
The final GRC is determined by reducing the intrinsic GRC through mitigations. Three types of mitigations, M1, M2, and M3, are possible as shown in Table~\ref{table:mitigations_grc}. The applicant identifies the possible mitigations and applies them. The reduction in GRC for each mitigation is evaluated based on the level of integrity (safety gain) and level of assurance (guarantee). M1 mitigations deal with the strategies used for lowering the density of people in the area, including controlling ground area, time of operation, and containment measures. M2 mitigations deal with the reduction of the impact of the UAV on the ground after a complete loss of control leading to a crash, including  parachutes, extensive simulations, or flight tests that show a single malfunction of a component doesn't lead to loss of control. M3 mitigations deal with the design of an ERP for addressing the situations in case of loss of control that cannot be handled by contingency procedures and lead to grave and imminent danger of fatalities. ERP should be documented in the form of a manual and cover plans that limit escalating effects of a crash and conditions for alerting air traffic management. The final GRC is determined based on the reductions accumulated from all three mitigations.

\begin{table*}[!ht]
\caption{Mitigations for final GRC determination.}
\centering
\begin{tabular}[!h]{ c c c c c }
\hline\hline
& & \multicolumn{3}{c}{\bf Robustness}\\
\hline
{\bf Mitigation Sequence} & {Mitigations for ground risk} & {Low/None} & {Medium} & {High}\\
\hline
1 & M1 — Strategic mitigations for ground risk & 0:None & -2 & -4\\
&  & -1:Low &  & \\
2 & M2 — Effects of ground impact are reduced & 0 & -1 & -2\\
3 & M3 — An ERP is in & 1 & 0 & -1\\
& place, the UAV operator is validated and effective &  &  & \\
\hline\hline
\end{tabular}
\label{table:mitigations_grc}
\end{table*}

\textbf{4) Initial Air risk class (ARC):}
 ARC relates to the intrinsic risk of a mid-air collision with another flying vehicle. ARC is dependent on the generalized density rating and the corresponding Airspace Encounter Category (AEC). The operational volume defined in ConOps forms the basis for determining the AEC. The initial ARC is determined based on the flying altitude, type of airspace, $i.e.$, controlled or uncontrolled, type of environment, $i.e.$, airport/heliport or non-airport/non-heliport, type of area, $i.e.$, urban or rural, and type of airspace, $i.e.$, typical or atypical, shown in Table~\ref{table:initial_arc}.

\begin{table*}[!ht]
\caption{Initial air risk class assessment.}
\centering
\begin{tabular}[!h]{ c c c c }
\hline\hline
\multicolumn{4}{c}{\bf Operational environment, AEC and ARC}\\
\hline
{\bf Operations in} & {\bf Initial generalized} & {\bf Corresponding AEC} & {\bf Initial ARC}\\
& {\bf density rating} &  & \\
\hline
{\bf Airport/heliport environment} &  &  & \\
\hline
OPS in an airport/heliport environment in & 5 & AEC 1 & ARC-d\\
class B, C or D airspace &  &  & \\
OPS in an airport/heliport environment in & 3 & AEC 6 & ARC-c\\
class E airspace or in class F or G &  &  & \\
\hline
{\bf Operations above 400 ft AGL but below flight level 600} &  &  & \\
\hline
OPS $>$ 400 ft AGL but $<$ FL 600 in a Mode-S & 5 & AEC 2 & ARC-d\\
Veil or transponder mandatory zone (TMZ) &  &  & \\
OPS $>$ 400 ft AGL but $<$ FL 600 in controlled & 5 & AEC 3 & ARC-d\\
airspace &  &  & \\
OPS $>$ 400 ft AGL but $<$ FL 600 in & 3 & AEC 4 & ARC-c\\
uncontrolled airspace over an urban area &  &  & \\
OPS $>$ 400 ft AGL but $<$ FL 600 in & 3 & AEC 5 & ARC-c\\
uncontrolled airspace over a rural area &  &  & \\
\hline
{\bf Operations below 400 ft AGL} &  &  & \\
\hline
OPS $<$ 400 ft AGL in a Mode-S Veil or TMZ & 3 & AEC 7 & ARC-c\\
OPS $<$ 400 ft AGL in controlled airspace & 3 & AEC 8 & ARC-c\\
OPS $<$ 400 ft AGL in uncontrolled airspace & 2 & AEC 9 & ARC-c\\
over an urban area &  &  & \\
OPS $<$ 400 ft AGL in uncontrolled airspace & 1 & AEC 10 & ARC-b\\
over a rural area &  &  & \\
\hline
{\bf Operations above flight level 600} &  &  & \\
\hline
OPS $>$ FL 600 & 1 & AEC 11 & ARC-b\\
OPS in atypical/segregated airspace & 1 & AEC 12 & ARC-a\\
\hline\hline
\end{tabular}
\label{table:initial_arc}
\end{table*}

\textbf{5) Final ARC:}
UAV operational volume may have a different collision risk from the one that the initial ARC assigned. If the generalised initial ARC assigned is too high for the condition in the local operational volume, then the ARC is to be reduced. This can be categorised into two different approaches, namely, strategic mitigation by the application of operational restrictions controlled by the operator, and strategic mitigation by the application of common structures and rules, which cannot be controlled by the operator. 
Operational restrictions are intended to mitigate the risk of a collision prior to take-off, including bounding the geographical and
operational time frame, and limiting the exposure time. Common flight rules mitigation is accomplished by setting a common set of rules which all airspace users must comply with, $e.g.$, coordination schemes, conspicuity requirements, and cooperative identification system. Also, common airspace structure mitigation is accomplished by controlling the airspace infrastructure through physical characteristics, procedures, and techniques. 

\textbf{6) Tactical Mitigation Performance Requirement (TMPR) and robustness levels:}
 TMPR is to mitigate any residual risk of a mid-air collision for airspace safety objectives. It is in the form of either ‘see and avoid’, $i.e.$, operations under VLOS, or additional systems, $e.g.$, DAA. VLOS is acceptable for collision risk for all ARC levels and no TMPR is required. EVLOS may require additional systems. The communication latency between the remote pilot and the observers should be less than 15 seconds. Therefore, in both cases, de-confliction scheme and phraseology is required. For operations other than VLOS, the applicant uses the residual ARC to determine required level of TMPR robustness. 

\textbf{7) Specific Assurance and Integrity Level (SAIL):}
SAIL represents the level of confidence that operations remain under control. The parameter combines the ground and air risk analyses into a single class. SAIL is determined based on the final GRC and residual ARC using Table~\ref{table:sail_table}. 

\begin{table}[!ht]
\caption{SAIL Determination.}
\centering
\begin{tabular}[!h]{ c c c c c }
\hline\hline
\multicolumn{5}{c}{\bf SAIL determination}\\
\hline
& \multicolumn{4}{c}{\bf Residual ARC}\\
\hline
{\bf Final GRC} & {\bf a} & {\bf b} & {\bf c} & {\bf d}\\
\hline
$\leq$ 2 & 1 & 2 & 4 & 6\\
3 & 2 & 2 & 4 & 6\\
4 & 3 & 3 & 4 & 6\\
5 & 4 & 4 & 4 & 6\\
6 & 5 & 5 & 5 & 6\\
7 & 6 & 6 & 6 & 6\\
$\geq$ 7 &  \multicolumn{4}{c}{Category C Operation}\\
\hline\hline
\end{tabular}
\label{table:sail_table}
\end{table}

\textbf{8) Operational Safety Objectives (OSOs):}
The SAIL assigned for the proposed ConOps determines what OSOs should be complied with and the associated level of robustness. The OSOs are classified into four categories based on the type of threat they mitigate: 1) Technical issue with the UAV, 2) Deterioration of external systems supporting UAV operations, 3) Human error, and 4) Adverse operating conditions. For each OSO, the level of integrity and assurance must be provided. 

\textbf{9) Adjacent area/airspace considerations:}
Adjacent area and airspace considerations cover infringement by the UAV and the risk posed due to loss of control of the operation. The extent of adjacent airspace is defined starting from the border of operational volume. Once the adjacent airspace is determined, the ground area should be fully researched to know the kinds of spaces, like buildings, parks, roads, etc. The nominal safety requirements should make sure that no probable failure of any component onboard the UAV or external supporting system leads to an operation outside the operational volume. In some cases, $e.g.$, assemblies of people or ARC-d category, an enhanced containment system is required.

\textbf{10) Comprehensive safety portfolio:}
The comprehensive safety portfolio summarizes the risk analysis and the mitigations for intrinsic GRC, initial ARC, adjacent area and airspace, and any additional requirements.

\section{SORA Application} \label{sec:aplication}
We are the Automation and Robotics Research group at the Interdisciplinary Centre for Security, Reliability, and Trust, University of Luxembourg. We conduct research on multirotor aerial robots involving multiple research fields such as Perception, State Estimation, Situation Awareness, Motion Control, and Multi-Robot Coordination, on our Kirchberg campus \cite{castillo2020real,habibi2022safe,sanchez2016aerostack}. Considering our outdoor UAV experimental operations, we have prepared the SORA application and recently obtained flight authorization. On this basis, we briefly present our application, addressing all the steps, as follows.

\textbf{1)} In ConOps, we provided the relevant technical, operational, and system information. In the first part, we presented a general overview of our organization, safety management system and measures, and our group's research activities. Then we explained in detail the training and maintenance process, crew roles, design and installation forms. 
Our operations are VLOS data acquisition, navigation and guidance improvements, and controller modification (manual or automatic), with identified general conditions and limitations for them, in a controlled populated area. We developed an operational manual, contingency and emergency, and occurrence reporting procedures, ERP, and training manual.

In the second part, we provided necessary and sufficient technical information about our UAVs and their supporting systems. We adopted a simple control segment comprising an autopilot, like Pixhawk, supported by onboard IMU sensors and GPS for navigation, and a remote pilot station for manual control. We do not have any DAA systems. 
We proposed to use the geo-caging feature in autopilot to confine the UAV to the maximum altitude, $i.e.$, 30 $m$, and the boundary of the region of flight. Additionally, visual observers onsite monitor and notify the remote pilot about UAV leaving the confinement zone for recovery action. Due to our proximity to an airport, we also proposed to have an enhanced containment system comprising a tether of length 30 $m$ suspended from the UAV and fixed to a heavy deadweight. If the tether breaks and all the containment systems fail, Flight Termination System (FTS) on the remote controller is activated.

%We adopted a C2 link architecture, as illustrated in Fig.~\ref{fig:c2_link_architecture}. 
A Ground Control Station (GCS) and the remote controller are used for C2 links. High-level mission planning is carried out in the GCS. The GCS operator and remote pilot continuouslys monitor their respective C2 link statuses and notifies the other in case of any link degradation or loss for taking necessary actions to enhance link quality or regain connection.
%\begin{figure}[!htpb]
% \centering
% \includegraphics[width=90mm]{c2_link_architecture.png}
% \caption{Command and control link architecture.}
% \label{fig:c2_link_architecture}
%\end{figure}
At the end of this part, we identified the most probable single failure modes, including malfunctioning of an autopilot component, one of the C2 links loss, GPS loss, battery malfunction, and propeller or airframe damage, and developed corresponding contingency, emergency, and recovery actions.

\textbf{2)}
The total, operational, and flight geography areas are determined, based on maximum tether length, $i.e.$, 30 $m$. There is no need for a containment zone, due to the tether. The flight geography is a circular region of a 30 $m$ radius around the takeoff points, at which the geo cage is set. A 5 $m$ region outside the flight geography is chosen as the buffer zone. We have identified the operational footprint in the shape of a polygon. Then we reduced the total length to 35  $m$, $i.e.$, the length of the tether (30 $m$) plus 5  $m$ of the buffer zone, to obtain the area, in which the tether is fixed to the ground, as illustrated in Fig. \ref{fig:flowchart}. There are two observers one on the ground and one on the top of the building, making sure that no one intrudes into the flight zone (flight geography plus buffer zone).
We temporarily block the building entrance points if any in the flight region during the operations. We close the buffer zone with barricades to have a Controlled Ground Area (CGA).
As per the specifications of UAVs and taking the maximum weight into account, the terminal velocity and maximum kinetic energy are computed. Then, our intrinsic GRC is determined as 2, according to Table \ref{table:intrinsic_grc}.

\textbf{3)}
We proposed mitigations and carefully investigated their robustness, to reduce the intrinsic risk of a person being struck by the UAV.
Since our operations are tethered in a CGA, the M1 in Table \ref{table:mitigations_grc} is not applicable. 
For M2, we selected “low”, since our UAVs are electrical ones, with an FTS. In case of FTS failures, propeller guards  forbid any adverse effect.
For the assurance level of M2, we selected “low”. 
For M3, we selected “medium”, as we have defined ERP. For the assurance level, we selected “medium”,  guaranteed by pilot training procedure, staff training record and pre-flight checklists. Considering M2-M3 mitigations, the GRC is not changed, considering Table \ref{table:mitigations_grc}.

\textbf{4)}
Our operations are below FL600, in controlled airspace in airport/heliport environment. Therefore, our operation falls in the ARC-d category. The corresponding AEC and density rating, according to Table \ref{table:initial_arc} are AEC-1 and 5, respectively. 

\textbf{5)}
As per the rules stipulated by Regulation (EU) No 923/2012, no lowering of the ARC by common structures and rules is allowed, as those mitigations were already accounted for while determining the intrinsic ARC.
The geographical boundary restrictions are enforced by using the tether as well as the geo-caging. Additionally, there are buffer zones in both horizontal and vertical directions for any unexpected space violations. In any direction, a buffer space of 5 $m$ is used. Any manned aircraft cannot fly below 150 $m$. So, a manned aircraft cannot enter our flight zone. Moreover, we proposed to notify the airport to lower the risk of encountering a manned aircraft. This reduced our ARC to ARC-b.

\textbf{6)}
Since our operations are all within VLOS, it is considered to be acceptable tactical mitigation for collision risk. In the case of potential incoming traffic, we follow the see and avoid strategy, as explained in the de-confliction scheme, including the phraseology, documented in the operational manual.

\textbf{7)}
Based on Table \ref{table:sail_table}, our SAIL is determined as, II. 

\textbf{8)}
As SAIL category was II, the level of integrity and assurances for the required OSOs, are explained as follows.
\begin{itemize} [leftmargin=*]
    \item Ensured that we are a competent UAV operator.
    \item Created design and installation appraisal form to record the maintenance operations.
    \item Provided our command, control, and communication link characteristics and monitoring process.
    \item Ensured UAV inspection consistency and included the inspection process in our training manual.
    \item Created an operational manual, explaining operational procedures, flight planning, unexpected adverse conditions, normal, contingency, emergency, and occurrence reporting procedures. Added pre/post-flight inspections and environmental conditions evaluation to the checklists.
    \item Included the crew training process in the training manual.
    \item Safety features to avoid operations outside flight zone.
    \item Clarified how external supporting systems, like U-Space, are not ready and applicable for our flight operations.
    \item Documented procedures for multi-crew coordination in the training manual.
    \item Created policies for remote crew fitness and resting times, as per the pre-flight checklist and logbook.
    \item Evaluated human factors for appropriate human-machine interfaces in the training manual.
    \item Included the assessment of the meteorological conditions in the pre-flight checklist.
\end{itemize}

\textbf{9)}
We considered the area covered by our UAV flying at a maximum speed flight speed of 2 $m/s$ for three minutes from the boundary of the ground buffer, as the adjacent area, where we have residential buildings, roads, a tram line, schools, etc. In case of an emergency, we contact the Campus Security and Safety Team. The adjacent airspace is ARC-d, requiring an enhanced containment system. Due to the use of tether, M1 mitigation is not applicable. Also, the probability of failure is demonstrated to be low. The probability of our UAVs leaving the operational volume is less than 10-4/FH due to the use of tethering, geo cage and flight termination system. The specification of the tethering system, to support the enhanced containment, is given in ConOps. Development errors of software and airborne electronic hardware (AEH) could not lead to operations outside the ground risk buffer.

\textbf{10)}
We created a comprehensive safety portfolio, summarizing all of previous steps and identified additional forms.

\section{Discussion on the challenges and potential solutions} \label{discussion}
Considering our experience with the SORA application, we have identified many challenges and issues. The most challenging issue is the ConOps development, as an iterative process, $i.e.$, additional mitigation and limitations might be identified later on, requiring additional associated technical details and procedures in the ConOps, which are not foreseen initially in the instructions. So, it can be a time-consuming process, with significant revisions or even changes in the operation. Therefore, this should culminate in a comprehensive ConOps that fully and accurately describes the operation as envisioned. Also, there is a requirement for an expert, with technical knowledge, to address the level of integrity and assurances, while there is no clear reference, nor examples to guide the applicant. Also, the guidelines are vague in most of parts and the workflow is difficult to interpret. 
More importantly, there are many checklists, manuals, and procedures, required to be prepared, which are not explicitly specified in the guidelines. However, preparing these documents at first avoids iterative and confusing procedures and delays. On the other hand, since there are no certified UAVs that satisfy the safety objectives, it is challenging to straightforwardly address safety mitigations. Also, the third-party validation is required for high SAIL categories.
\subsection{Additional technical documents}\label{additional}
To overcome the above-mentioned challenges, here, we propose a comprehensive list of additional technical documents, to be prepared preliminarily, which facilitate the preparation of the application, since we foresee and implement all the required procedural documents, using which the iterative process is avoided. For this, we identify and explain these additional technical documents and their content, to be prepared by the applicants initially, which is taken into account in the proceeding steps, as follows.

\begin{itemize} [leftmargin=*]
    \item Checklists:
    \begin{itemize} [leftmargin=*]
    \item Pre-flight UAV inspection and related considerations: Crew is familiar with the operational manual, design and installation appraisal, training manual, training record table, briefed on the mission plan with identified obstacles. The remote pilot is certified to fly. UAV is registered/labelled and insured. The tether (if required) is installed, with no sign of wearing, tearing and tied securely to dead weight and UAV. The region of flight is within the flight geography. Take-off, landing, return-to-home, geo-fence/geo-cage, altitude ceiling and emergency landing points are specified and loaded to GCS and UAV. Batteries are fully charged and securely mounted. No signs of damage in the airframe and all screws are tight. Propeller blades have no cracks, are not loose, and are deformed. Propulsion system mountings and payloads are secure. The communication link among RC, GCS and UAV is established. GPS is locked if being used. Confirm the phone number of the nearest Air Traffic Control facility in the event of an emergency. GSE is available (spare parts, such as propellers, airframe, batteries, walkie-talkies, internet devices and a first-aid kit). Collecting personal data is checked. GDPR procedures are implemented for data subjects’ awareness. The flight zone is secured. FTS is functional, and ground impact measures, $e.g.$ propeller guards, are in place. Airspace management authority is notified before operation.
    \item Pre-flight crew fitness: Crew is under no medication, visually not impaired, not fatigued, not intoxicated, physically, mentally and auditorily fit.
    \item Pre-flight environmental conditions: Environmental conditions within UAV's performance limits, are identified and checked before the operation.
    \item In-flight human error assessment: Remote pilot recovers out-of-control UAV. UAV status is constantly monitored. No obstacle is overlooked. GCS operator monitored UAV performance and conditions adequately. Communication protocols between the crew are not violated.
    \item Post-flight inspection and related considerations: UAV powered off and batteries disconnected, not overheated and cooled down for recharging, checked for any signs of damage. Flight duty cycles and maintenance reports are completed. The motors are not overheated. GDPR management (transferring, storing, limiting access, security, and minimization of personal data).
    \end{itemize}
    \item Design and installation appraisal: Logging on the date, UAV label, UAV configuration version, modification/maintenance (staff, added/removed/replaced components, modified total weight, configuration version), including, airframe, airframe material, motors, propellers, ESCs, flight controller, power module, remote controller, RF communication, telemetry, GPS, camera, onboard computer, additional sensors, autopilot software, Ubec, battery, total weight, manufacturer, description and redundant, independent, separate component.
    \item Logbook: Logging on the date, start and end time, UAV label, configuration version, battery labels, takeoff mass, payload, mission description, crew members, resting times, incidence, and comments.
    \item De-confliction scheme: Defining the phraseology for communicating about the adjacent air space, monitored continuously for potential incoming air traffic, and the corresponding actions.
    \item Staff Training Record: Recording/updating the staff training, including theoretical courses, UAV inspection and maintenance, practical training, crew resource management and coordination, contingency procedures, emergency procedures and ERP, GDPR procedures, and date of refresh training.  
    \item General Data Protection Regulation (GDPR) Procedures:
    Identifying the privacy risks of the operation, defining the role with respect to personal data collection and processing, data protection impact assessment, describing the measures to ensure data subjects are aware that their data may be collected, describing the measures to minimise the personal data collecting, the procedure established to store the personal data and limiting access, the measures taken to ensure that data subjects can exercise their right to access, correction, objection, and erasure.   
    \item Operational manual:
    \begin{itemize} [leftmargin=*]
        \item Flight Planning: Determining site map, region of flight, obstacles, mission objectives, ground reference points, flight parameters (take-off \& landing spots, flying height, speed, geo-cage, wind), GSE and crew roles.
        \item Pre- and post-flight Inspection: Verifying Checklist items.
        \item Environmental Conditions Evaluation: Obtaining the weather condition information from the official meteorological website. Verifying checklist items.
        \item Normal Operations Procedures: Pre-flight routines: Verifying pre-flight UAV inspection, crew fitness, and environmental conditions evaluation. Flight planning, Mission briefing. In-flight routines: Continuous monitoring for human error assessment, UAV collision, De-confliction scheme, communication links, UAV confinement to flight geography, the battery level, and absence of obstacles for landing, landing and powering off. Post-flight routines: Filling post-flight checklist and logbook. 
        \item Contingency Procedures: Identifying unexpected adverse operating conditions and defining recovery plans.
        \item Emergency Procedures: Loss of communication, loss of control, fly away (loss of containment), pilot incapacitation, intrusion into flying area, failure of enhance containment system, fire or injury after the crash, with corresponding responses by each crew member.
        \item ERP: Serious personnel injury (calling ambulance or police), incidents with manned aircraft (calling corresponding authority), flying away (calling the airport), fire (calling fire department) or any other accident (calling the police, site security), with identified contact numbers are available, following Occurrence Reporting Plan.
        \item Occurrence Reporting Plan: Fatal/serious injury to a person, and incidence with a manned aircraft, details about the aircraft, crew, events, and a narrative, In addition, all the event-related flight logs are saved for sharing with competent authorities when needed for any investigation.
    \end{itemize}
    \item Training manual:
    \begin{itemize} [leftmargin=*]
        \item Theoretical and practical courses by the authority: Air safety, airspace restrictions, geo-awareness, UAV aviation regulations, airspace categorization, UAV classes, human performance limitations, risk awareness, planning and performing a flight, operational procedures, remote pilot responsibilities, pre/in/post flight operations and checklists, UAV components general knowledge, GDPR, UAV insurance and security, weather conditions, atmospheric circulations, with the corresponding resources.
        \item UAV Inspection and maintenance: Flight crew familiarity with the design and installation appraisal, UAV registration, labelling, and insurance, pre- and post-flight inspections and maintenance for battery, airframe, propeller blades, propulsion system, payload, UAV operations after landing and completion of maintenance report.
        \item Crew Resource Management and coordination: Crew familiarity with different roles and corresponding responsibilities in UAV operations, with communication means.
        \item Possible human errors pre-and in-flight: Crew is briefed prior to the operation about flying the UAV out of the flight geography, losing LOS throughout the mission, overlooking obstacles, inadequate monitoring UAV and conditions, and violation of communication protocols.
        \item Meteorological conditions assessment: The weather conditions, including the wind speed and direction, visibility, storm, rain, and snow forecasts.
        \item Contingency Procedures and Emergency Procedures and ERP: Crew should be trained for the contingency procedures and emergency procedures and ERP.
        \item GDPR procedures: Crew is be familiar with GDPR.    
    \end{itemize}
\end{itemize}

\subsection{Alternative workflow} \label{alternative}
The proposed alternative workflow is illustrated in Fig. \ref{fig:flowchart}, by implementing the additional documents, presented in Section \ref{additional}, into the original SORA methodology workflow, which addresses challenges by avoiding time-consuming iterations. It is worth noting that there is a possibility to choose SAIL category based on the applicant's available mitigations, and the applicant adapts the operation to satisfy the required ground and air risk classes. However, this approach is suitable for specific operations, not generic ones.
\subsection{Automation approach}
In addition to the proposed workflow in Section \ref{alternative}, here, we propose an envisioned automated alternative approach, in which we can encapsulate most of the steps into computer software, to be followed by the applicants. This approach is illustrated in Fig. \ref{fig:flowchart} and explained as follows.

Initially, Table \ref{table:intrinsic_grc}, can be programmed to let the user insert the location of the operation and specifications of the UAV to compute the initial GRC. Moreover, GRC mitigation options \ref{table:mitigations_grc} can be automatically given for the user to have the required criteria and choose possible mitigation to have the final GRC. The EU-level database is to be used where coordinates of the area of operation can be given, and the output would be initial ARC, taking into account Table \ref{table:initial_arc}. Similarly, the user can be given the mitigation means, examples and TMPR options, and upon selection, the final ARC can be determined. Then, SAIL can be automatically determined. Based on the chosen area of operation in the previous steps, adjacent area/airspace can be automatically determined and a need for enhanced containment is identified. At this point, the user should be given options for containment, which can result in changes to GRCs and ARCs, as well as ConOps or continue by addressing the containment requirements. Accordingly, the user is presented with the list of required OSOs, and associated robustness levels to be addressed. For each OSO, ideas and examples can be presented to the user to address the criteria. Finally, the program summarizes the application in the form of a safety portfolio. This automated workflow is illustrated in Fig. \ref{fig:flowchart}. 

\begin{remark}
 There will be some updates in the guidelines in SORA 2.5, $e.g.$, using U-space structures and rules, which might impose additional requirements for preparing the application. However, the proposed automated approach can still be used, taking into account the updated parts.
\end{remark}
% \begin{remark}
% It is highly recommended for applicants have a preliminary discussion with the regulatory body on their intended operations before proceeding with application preparation.
% \end{remark}

\section{Conclusions} \label{sec:conclusions}
In this paper, we carefully investigated the whole process to obtain outdoor flight authorization, addressing Specific Operations Risk Assessment (SORA) requirements. We highlighted the observed challenges, issues and pitfalls in the workflow. Accordingly, we proposed an alternative approach, following the same steps, to shorten the whole process, with a comprehensive list of preliminary procedures. We also suggested an automated workflow, which can be implemented as computer software. It is aimed to help UAV operators and the research community to eliminate the difficulty and delays in conducting intended operations.
\bibliographystyle{ieeetr}
\bibliography{references}

\begin{thebibliography}{10}

\bibitem{srivastava2020review}
S.~Srivastava, S.~Gupta, O.~Dikshit, and S.~Nair, ``A review of uav regulations
  and policies in india,'' {\em Proceedings of UASG 2019: Unmanned Aerial
  System in Geomatics 1}, pp.~315--325, 2020.

\bibitem{dronelaws}
``Drone laws by countries, states, cities,
  \hyperlink{https://drone-laws.com/}{https://drone-laws.com/},''

\bibitem{EASA}
``Specific operations risk assessment,
  \hyperlink{https://www.easa.europa.eu/en/domains/civil-drones-rpas/specific-category-civil-drones/specific-operations-risk-assessment-sora}{https://www.easa.europa.eu/en/domains/civil-drones-rpas/specific-category-civil-drones/specific-operations-risk-assessment-sora},''

\bibitem{usalaws}
``Drone laws in the united states of america,
  \hyperlink{https://uavcoach.com/drone-laws-in-united-states-of-america/}{https://uavcoach.com/drone-laws-in-united-states-of-america/},''

\bibitem{chinalaws}
``Civil aviation administration of china,
  \hyperlink{http://www.caac.gov.cn/index.html}{http://www.caac.gov.cn/index.html},''

\bibitem{canadalaws}
``Canadian aviation regulations,
  \hyperlink{https://tc.canada.ca/en/aviation/drone-safety/learn-rules-you-fly-your-drone/flying-your-drone-safely-legally}{https://tc.canada.ca/en/aviation/drone-safety/learn-rules-you-fly-your-drone/flying-your-drone-safely-legally},''

\bibitem{xu2020recent}
C.~Xu, X.~Liao, J.~Tan, H.~Ye, and H.~Lu, ``Recent research progress of
  unmanned aerial vehicle regulation policies and technologies in urban low
  altitude,'' {\em IEEE Access}, vol.~8, pp.~74175--74194, 2020.

\bibitem{stocker2017review}
C.~St{\"o}cker, R.~Bennett, F.~Nex, M.~Gerke, and J.~Zevenbergen, ``Review of
  the current state of uav regulations,'' {\em Remote sensing}, vol.~9, no.~5,
  p.~459, 2017.

\bibitem{castillo2020real}
M.~Castillo-Lopez, P.~Ludivig, S.~A. Sajadi-Alamdari, J.~L. Sanchez-Lopez,
  M.~A. Olivares-Mendez, and H.~Voos, ``A real-time approach for
  chance-constrained motion planning with dynamic obstacles,'' {\em IEEE Robot.
  Autom. Lett.}, vol.~5, no.~2, pp.~3620--3625, 2020.

\bibitem{habibi2022safe}
H.~Habibi, A.~Safaei, H.~Voos, M.~Darouach, and J.~L. Sanchez-Lopez, ``Safe
  navigation of a quadrotor uav with uncertain dynamics and guaranteed
  collision avoidance using barrier lyapunov function,'' {\em Aerosp. Sci.
  Technol.}, p.~108064, 2023.

\bibitem{sanchez2016aerostack}
J.~L. Sanchez-Lopez, R.~A.~S. Fern{\'a}ndez, H.~Bavle, C.~Sampedro, M.~Molina,
  J.~Pestana, and P.~Campoy, ``Aerostack: An architecture and open-source
  software framework for aerial robotics,'' in {\em International Conference on
  Unmanned Aircraft Systems (ICUAS)}, pp.~332--341, IEEE, 2016.

\end{thebibliography}
\end{document}